\documentclass[onecolumn,floatfix,superscriptaddress,a4paper,
               showpacs,showkeys,nofootinbib,preprint]{revtex4}

\usepackage{epsfig}
\usepackage{latexsym}
\usepackage{xspace}
\usepackage{hyperref}
\usepackage[latin2]{inputenc}
\usepackage{indentfirst}
\usepackage{enumerate}

\usepackage{color}

\usepackage{amsmath}
\usepackage{empheq}
\usepackage{amssymb}
\usepackage[english]{babel}
\usepackage{url}

\newcommand{\bs}{\boldsymbol}

\begin{document}

\title{Four types of phase transitions in interacting boson (meson) matter
at high temperatures}

\author{D. Anchishkin}
\affiliation{Bogolyubov Institute for Theoretical Physics, 03143 Kyiv, Ukraine}
\affiliation{Taras Shevchenko National University of Kyiv, 03127 Kyiv, Ukraine}
\affiliation{Frankfurt Institute for Advanced Studies, Ruth-Moufang-Strasse 1,
60438 Frankfurt am Main, Germany}

\author{V. Gnatovskyy}
\affiliation{Bogolyubov Institute for Theoretical Physics, 03143 Kyiv, Ukraine}

\author{D. Zhuravel}
\affiliation{Bogolyubov Institute for Theoretical Physics, 03143 Kyiv, Ukraine}

 \author{I. Mishustin}
 \affiliation{Frankfurt Institute for Advanced Studies, Ruth-Moufang-Strasse 1,
 60438 Frankfurt am Main, Germany}

 \author{H. St\"ocker}
 \affiliation{Frankfurt Institute for Advanced Studies,
 Ruth-Moufang-Strasse 1, 60438 Frankfurt am Main, Germany}
 \affiliation{Johann Wolfgang Goethe University, D-60438 Frankfurt am
 Main, Germany}

\begin{abstract}
The thermodynamics of the interacting system of relativistic particles
and antiparticles in the presence of a Bose-Einstein condensate was
investigated within the framework of the mean-field model.
It is assumed that the total isospin (charge) density is conserved.
We show that, depending on the strength of the interaction, a bosonic
particle-antiparticle system exhibits four types of phase transitions to the
condensate phase.
Three types correspond to the second-order phase transition, while one is
a first-order phase transition.
\end{abstract}

\maketitle

\section*{Introduction}

Interacting meson matter can arise in processes such as heavy-ion
collisions and recent "TeV" proton collisions. The processes where
many particles are created after collision can be observed and
obtained with significant statistics. Most of these particles are
pi-mesons. To understand the behavior of such exotic matter,
theoretical models are essential. This study investigates the
formation of Bose-Einstein condensates in a pion-antipion system or
any other particle-antiparticle bosonic system with interactions.
When studying Bose-Einstein condensation in a relativistic Bose gas
at high densities, two critical factors must be considered: 1. High
densities imply that the condensate states occupy a region of high
temperatures, where the density of thermal particles can no longer
be treated as a small fluctuation compared to the condensate
density. 2. The conservation law governing the number of particles
minus the number of antiparticles must be accounted for. Therefore,
any study of Bose-Einstein condensation in a relativistic Bose gas
must consider antiparticles. This approach was first discussed in
\cite{haber-1981}. A scalar model of a bosonic system forming a
Bose-Einstein condensate with conserved isospin (charge) was studied
in Refs.~\cite{haber-1981,kapusta-1981,haber-1982}.

In this paper, we generalize previous results, which were obtained
by introducing the thermodynamic mean-field model
\cite{Anch-Vovchenko}. Using this approach, Bose-Einstein
condensation was first studied for systems with zero isospin density
\cite{Anch-JPG46,anchishkin-4-2019}, and later for systems with
finite isospin density ($n_I = n^{(-)} - n^{(+)} > 0$ \footnote{Here
$n^{\mp}$ are the number densities of the $\pi^-$ and $\pi^+$
mesons, respectively.}) within the Canonical Ensemble
\cite{Anch-PRC105,universe-2023}. We calculated the temperature
characteristics of a non-ideal hot pion gas for various relations
between attractive and repulsive parameters. Building on recent
findings, this paper introduces a classification of phase
transitions at high temperatures in a two-component meson system
with strong interactions and fixed isospin density $n_I$. The phase
transitions leading to Bose-Einstein condensate formation are
classified into four types. It is important to note that this
classification is independent of the order of the phase transition,
which is typically addressed in statistical thermodynamics, such as
the Ehrenfest classification \cite{ehrenfest-1933,jaeger-1998}. Each
type of phase transition has its specific order.


\section{The First Type of Phase Transition: Formation
of a Bose Condensate}
\label{sec:first-type}

\begin{figure}
  \begin{center}
  \includegraphics[width=0.49\columnwidth]{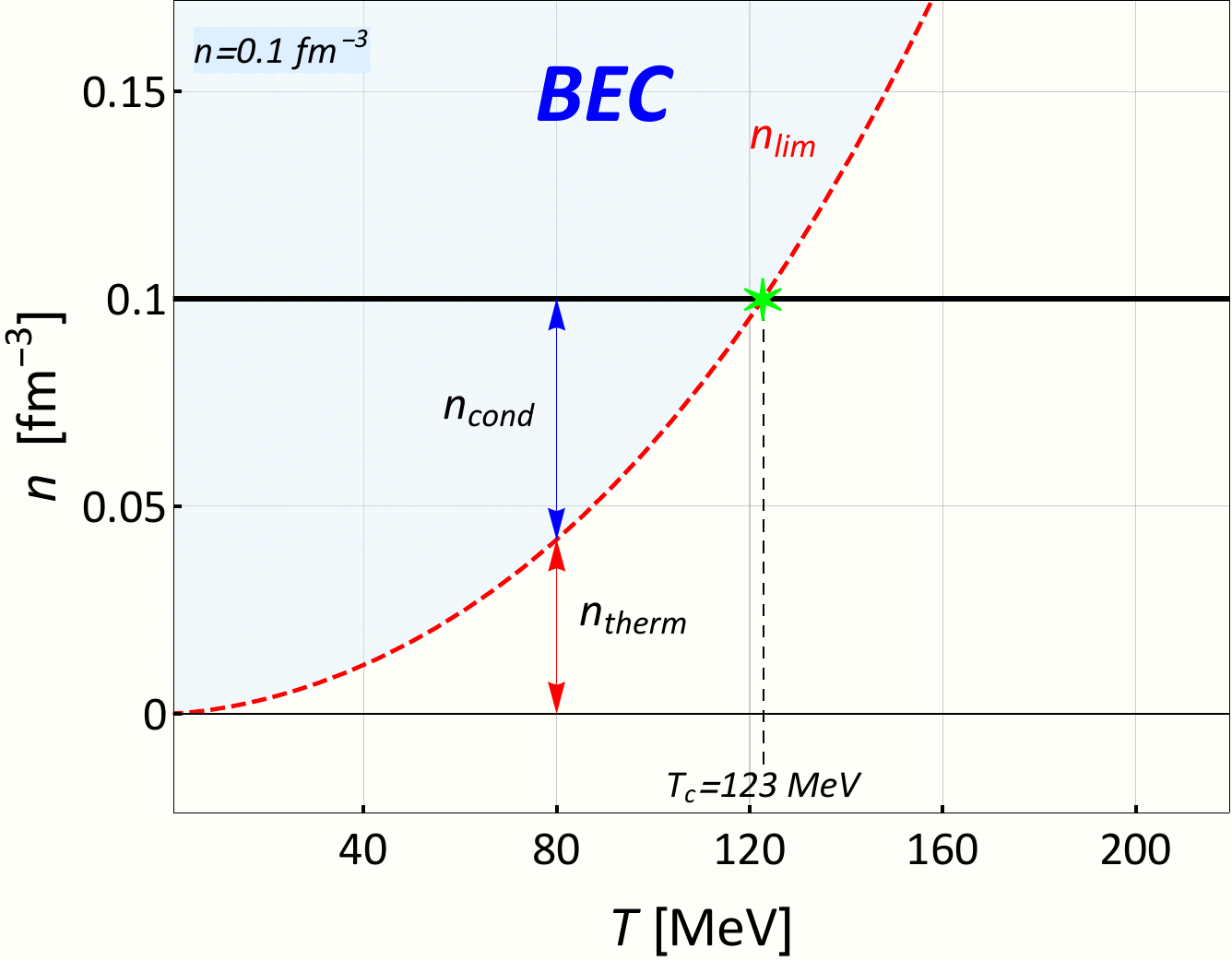}
  \includegraphics[width=0.49\columnwidth]{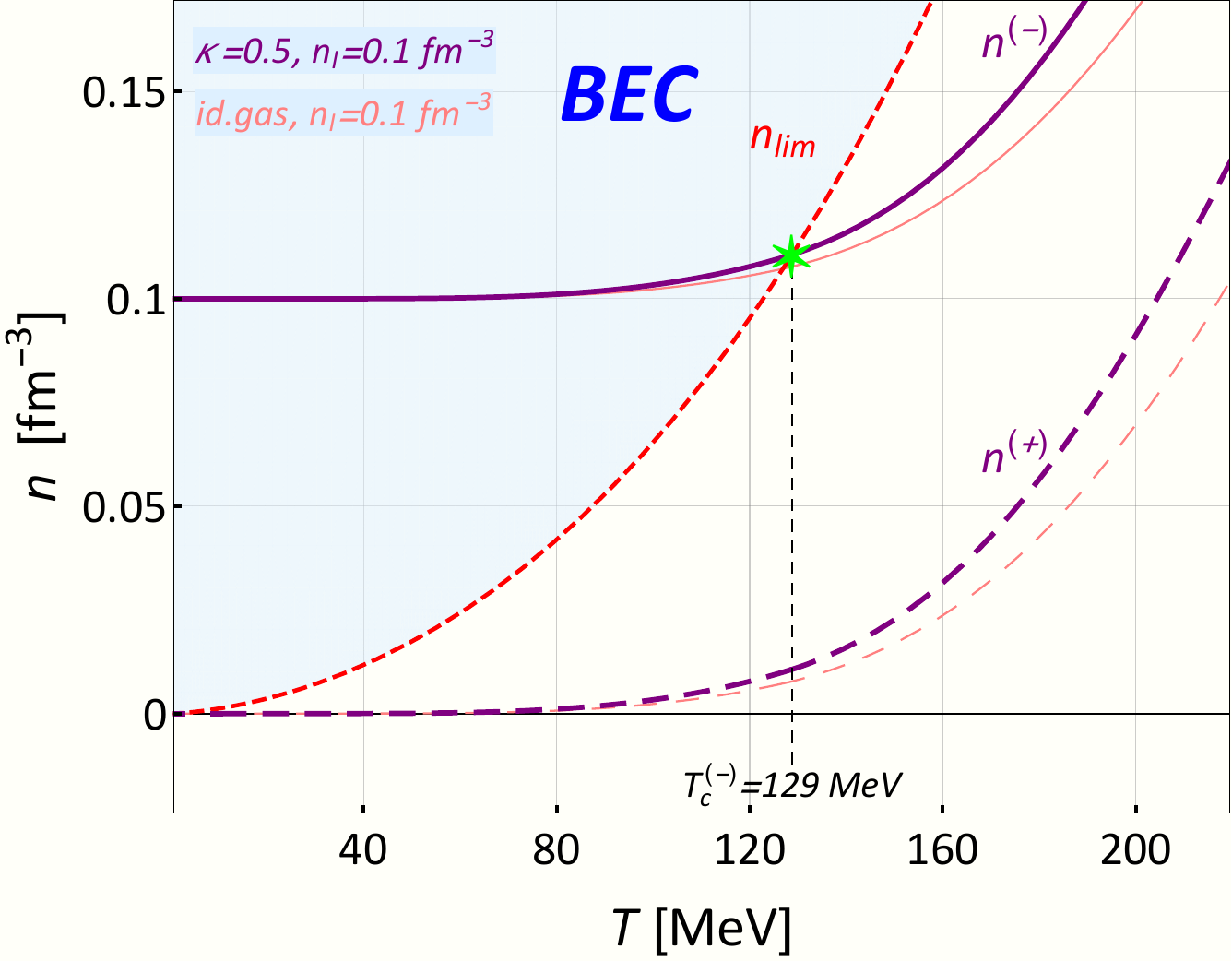}
  \caption{
  {\it Left panel:}
Dependence of the particle number density on temperature in an ideal
single-component bosonic gas at a constant particle density
$n = 0.1 \,\mathrm{fm}^{-3}$ with the corresponding critical temperature
$T_{\rm c} = 123$~MeV.
  {\it Right panel:}
Particle number densities $n^{(+)}$, $n^{(-)}$ on temperature at a constant
isospin density $n_I = 0.1 \,\mathrm{fm}^{-3}$ for an ideal gas (thin lines)
and an interacting pion gas in the mean-field model with an attraction
parameter $\kappa = 0.5$ (bold lines).
}
  \label{fig:ideal-gas}
  \end{center}
\end{figure}

We begin with a textbook example: the condensation of particles in a
single-component ideal bosonic gas.
Consider a boson system with constant particle-number density
in a box of fixed volume.
As the temperature decreases, the system undergoes a phase transition to the
condensate phase at a critical temperature, $T_{\rm c}$.
Below $T_{\rm c}$, the system is a mixture of condensed and thermal (kinetic)
particles.
In Fig~\ref{fig:ideal-gas} (left panel), the particle-number density
$n =$~const is shown as a horizontal solid line
(specifically, $n = 0.1$~fm$^{-3}$).
The red dashed line, marked $n_{\rm lim}$, is defined as:
\begin{equation}
n_{\rm lim}(T)  \,=\,  \int \frac{d^3k}{(2\pi)^3}\,
f_{_{\rm BE}}\big(\omega_k,\mu\big)\Big|_{\mu = m} \,,
\quad {\rm where} \quad
f_{_{\rm BE}}\big(E,\mu\big) \,
=\, \left[ \exp{ \left( \frac{E - \mu}{T} \right)}  - 1\right]^{-1} \,.
\label{eq:nlim-id}
\end{equation}
Here $f_{_{\rm BE}}\big(\omega_k,\mu\big)$ is the Bose-Einstein distribution
function with $\omega_k = \sqrt{m^2 + \bs k^2}$.
In Fig.~\ref{fig:ideal-gas}, the line $n_{\rm lim}$ divides the entire plane into
two parts:
A) The lower part represents the thermal or kinetic states of the particles;
B) The shaded part or upper part represents the states, where in addition to
the thermal particles there are the condensed particles as well.
The dependence $n_{\rm lim}(T)$ represents the maximum number of thermal particles
at the temperatures below $T_{\rm c}$ since the chemical potential reaches its
maximum allowed value, equal to the particle's mass, $\mu = m$.
The intersection of the lines $n =$~const and  $n_{\rm lim}$ determines the
critical temperature $T_{\rm c}$, marked as a bright star in
Fig~\ref{fig:ideal-gas} on the left panel.
As shown in \cite{universe-2023,Anch-PRC105}, at temperature $T = T_{\rm c}$
the energy density varies smoothly, while the heat capacity is continuous
but exhibits a discontinuity in its derivative, indicating a second-order
(continuous) phase transition.

Within the Canonical Ensemble, for a given value of the number density $n$
at temperatures $T \ge T_{\rm c}$, equation
$n = \int (d^3k)/(2\pi)^3 f_{_{\rm BE}}(\omega_k,\mu)$ determines the
dependence $\mu(T,n)$.
However, at temperatures $T < T_{\rm c}$, the chemical potential takes on
a maximum value equal to the particle's mass.
The equation becomes $n = n_{\rm cond}(T) + n_{\rm lim}(T)$, which determines
the number density of condensed particles $n_{\rm cond}(T)$.
The main thing to note is that the density of the condensate, which is an
order parameter, exists only in the temperature range $0 \le T < T_{\rm c}$
with zero density at the critical point $n_{\rm cond}(T_{\rm c}) = 0$, and
has a maximum value $n_{\rm cond}(0) = n$ at zero temperature, just like
the magnetization in ferromagnetic.
This behavior characterizes the first type of phase transition in bosonic
systems.

Next, we consider relativistic bosonic systems of particles and antiparticles
that have the same property of the appearance of condensate in the temperature
interval $0 \le T < T_{\rm c}$.
{\it The first example} is a relativistic ideal gas of bosons with a conserved
isospin density, i.e. $n_I = n^{(-)} - n^{(+)} =$~const, where $n^{(-)}$ is
the number density of negative particles, and $n^{(+)}$ is the number density of
positive antiparticles.
In this case, the Euler relation is $\varepsilon + p = sT + \mu_I n_I$.
The temperature dependencies of these densities are shown in
Fig.~\ref{fig:ideal-gas} in the right panel by thin lines.
In such systems, only one component (e.g., negatively charged particles) can
form a condensate phase provided that the necessary condition $m - \mu_I = 0$
is fulfilled.
One can argue that positively charged antiparticles can form a condensate
simultaneously with particles.
However, it should be noted that antiparticles have a chemical potential of
the opposite sign compared to particles; the necessary condition for the
formation of a condensate for antiparticles is $m + \mu_I = 0$.
These two conditions lead to the
equalities $\mu_I = 0$ and $m = 0$, which is impossible.
Thus, the necessary conditions for the simultaneous condensation of particles
and antiparticles cannot be fulfilled.
Only the component $n^{(-)}$ that carries the full isospin charge at zero
temperature, i.e. $n^{(-)}(T = 0) = n_I$, can form a condensate.
Due to this, the critical curve $n_{\rm lim}(T)$ is associated with the
$\pi^-$ mesons, as shown in Fig.~\ref{fig:ideal-gas} in the right panel.

{\it The second example} is the interacting particle-antiparticle system, where
the condensate exists only in the temperature range $0 \le T < T_{\rm c}$.

At high temperatures, thermally produced particles have a fairly high density.
Under such conditions, the interaction effects become important.
Thermodynamic properties of a system of an interacting boson particles and
antiparticles at high temperatures we study within the framework of the
thermodynamically consistent mean-field model
\cite{Anch-Vovchenko,Anch-JPG46,anchishkin-4-2019,universe-2023}.
The mean field contains both attractive and repulsive terms.
The phase structure of the bosonic system is regulated by the scaling
dimensionless parameter $\kappa$, which is the ratio between the attraction
and repulsion parameters
(for more details, see \cite{Anch-JPG46,universe-2023,Anch-PRC105}).
By increasing $\kappa$, we increase the intensity of attraction between
particles, while fixing the repulsion parameter.
Having overcome the critical value $\kappa = 1$, the system can enter the
regime of "strong" (supercritical) interaction, that is, when $\kappa > 1$.
Each regime exhibits its own phase structure with different phase transitions.

For the case of "weak" attraction, e.g. $\kappa = 0.5$, the situation is very
similar to an ideal gas of particles and antiparticles, as can be seen in the
right panel of Fig.~\ref{fig:ideal-gas}, where the particle and antiparticle
densities are shown by black solid and black dashed lines, respectively
(the thin lines correspond to the ideal gas).
Similar to an ideal gas, only $\pi^-$ mesons undergo a phase transition to the
condensed phase at $T = T_{\rm c}$, and the critical curve $n_{\rm lim}(T)$
is also associated only with particles.
The temperature $T_{\rm c}$ indicates a phase transition of the second order,
where the heat capacity is a continuous function with a discontinuity of the
derivative at this point.
Nevertheless, the dependence of the energy density on temperature is a smooth
function at $T_{\rm c}$ \cite{universe-2023,Anch-PRC105}.

So, in this section we have considered the {\it first type} of phase
transition - Bose condensate is formed only in the temperature range
$0 \le T < T_{\rm c}$, where $T_{\rm c}$ is the critical temperature of
the second-order phase transition in all three cases: a single-component
ideal gas, an ideal gas of particles and antiparticles, and an interacting
gas of particles and antiparticles with a moderate or weak intensity of
attraction between particles.


\section{The second type: Multiple formation of a Bose condensate }
\label{sec:second-type}

\begin{figure}
\centering
\includegraphics[width=0.44\textwidth]{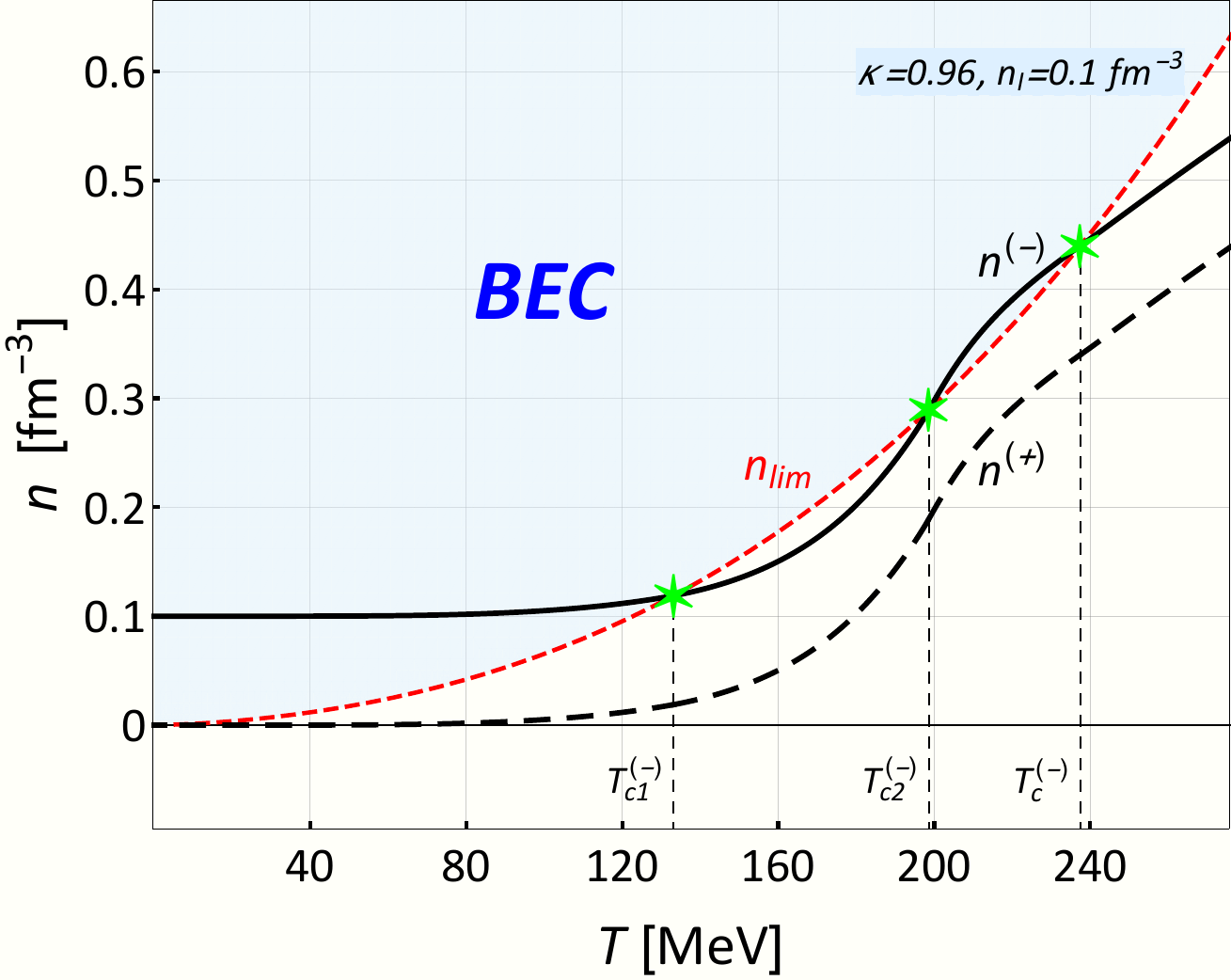}
\hspace{1mm}
\includegraphics[width=0.49\textwidth]{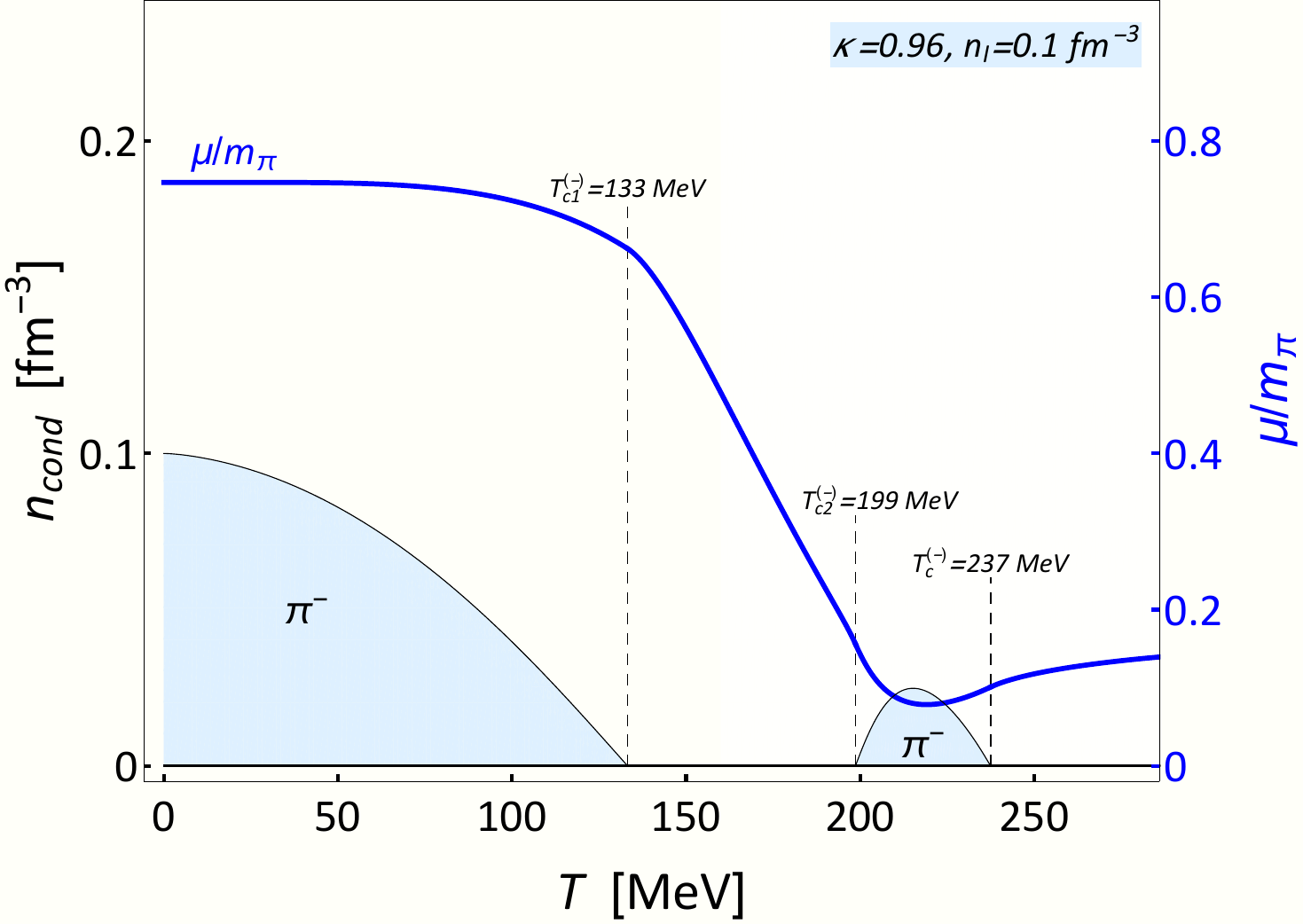}
\caption{
{\it Left panel:}
The particle-number densities $n^{(+)}$ , $n^{(-)}$ vs. temperature for
the interacting $\pi^+$-$\pi^-$ gas in the mean-field model at $\kappa = 0.96$
and isospin density $n_I = 0.1\, \mathrm{fm}^{-3}$.
The temperatures $T_{\rm c}^{(-)}$, $T_{\rm c1}^{(-)}$ and $T_{\rm c2}^{(-)}$
indicate the multiple phase transitions of the second order.
{\it Right panel:}
Dependence of the density of the $\pi^-$ meson condensate (shaded areas) on
temperature and the chemical potential (right Y-axis) on temperature for the
same boson system and the same conditions as in the left panel.
}
\label{fig:particles-antiparticles-kappa096_01}
\end{figure}
We still investigate the system of particles and antiparticles at a constant
isospin density.
Let us increase the intensity of attraction between particles and consider
the attraction parameter approaches the critical value, $\kappa \lesssim 1$.
In this case, the dependence of the density of $\pi^{-}$ mesons is shown in the
left panel of Fig.~\ref{fig:particles-antiparticles-kappa096_01} by a black
solid line.
Each point marked with an asterisk on the graph corresponds to a second-order
phase transition, where the $\pi^{-}$ component of the system undergoes multiple
phase transitions.
Indeed, by calculating the dependencies of the heat capacity and energy density
on temperature, it can be verified that each intersection of the critical curve
$n_{\rm lim}(T)$ with the curve $n^{(-)}( T)$ is a second-order phase transition,
see the left and right panels of Fig.~\ref{fig:heat-capacity-kappa1-01}.
In this case, the system undergoes multiple phase transitions with numerous Bose
condensate formations, shown in the right panel of
Fig.~\ref{fig:particles-antiparticles-kappa096_01} as the blue-shaded regions.
We call such multiple phase transitions with {\it multiple Bose condensate
formation} the second type of phase transitions.


\section{The third type: Meta phase transitions of the second order}
\label{sec:third-type}

In the previous section we investigate some peculiarities in the interacting
bosonic system that are due to behavior of $\pi^-$ component - the phenomenon
of the multiple phase transition.
Let us see what happens when we increase the attraction parameter and it equals
to the critical value $\kappa = 1$.
%
\begin{figure}
\centering            
\includegraphics[width=0.475\columnwidth]{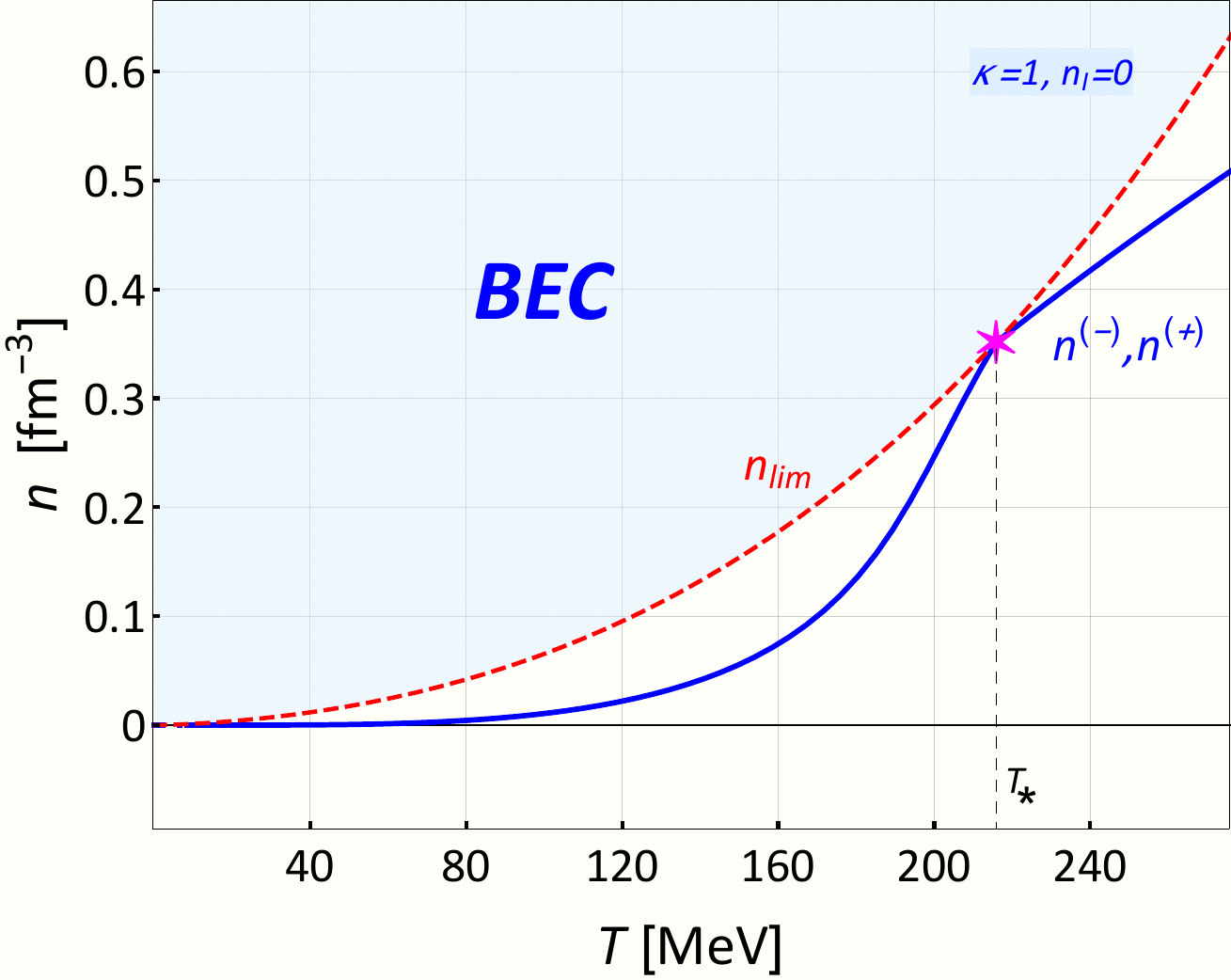}
\hspace{3mm}
\includegraphics[width=0.46\columnwidth]{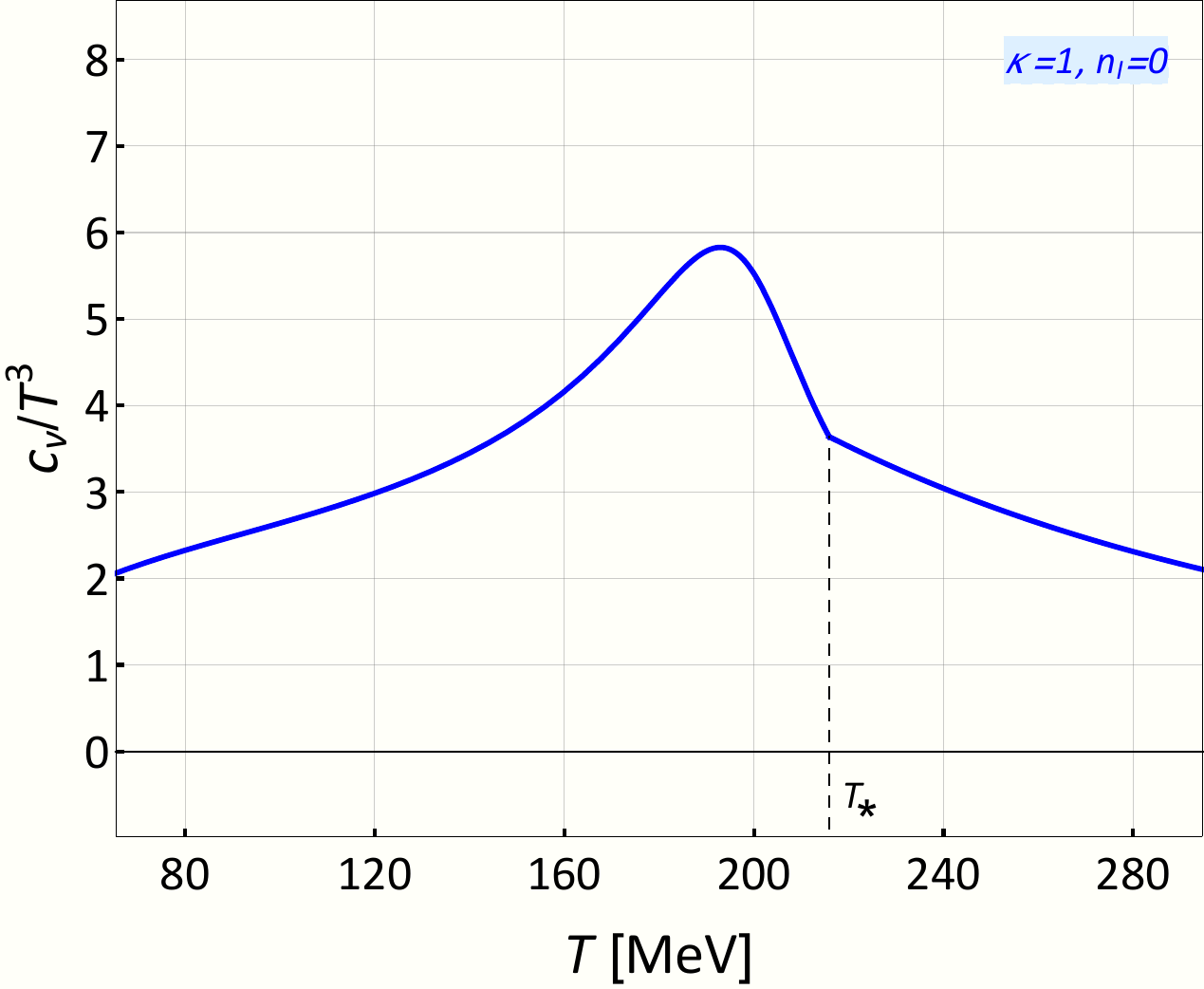}
\caption{ {\it Left panel:}
The particle-number densities $n^{(-)}$, $n^{(+)}$ vs.
temperature for the interacting $\pi^-$-$\pi^+$ gas in the mean-field model
at $\kappa = 1$ and isospin density $n_I = 0$.
The temperature $T_{*}$ indicates the meta phase transition of the second order.
{\it Right panel:} Heat capacity vs. temperature for the same boson system and the same conditions
as in the left panel.
}
\label{fig:meta-phase-trans}
\end{figure}
%
First, we consider a neutral system, i.e. $n_I = 0$, with $\kappa = 1$.
In this case, the densities of $\pi^-$ and $\pi^+$ mesons are equal to each
other, $n^{(-)}(T) = n^{(+)}(T)$.
The self-consistent solution for these particle-number densities is shown
in the left panel of Fig.~\ref{fig:meta-phase-trans} by the blue solid line
that touches the critical curve $n_{\rm lim}$ at temperature $T_*$.
As we can see in the right panel, the temperature dependence of the heat
capacity is a continuous function, but has a break at temperature $T_*$.
The energy density is a smooth function at this point \cite{Anch-JPG46}.
So, by all formal signs, this is a second-order phase transition.
Unlike standard phase transitions, this transition does not lead to the
formation of a condensate, as can be seen in the left panel
of Fig.~\ref{fig:meta-phase-trans}.
That is why we call it a {\it meta phase transition of the second order}.

\begin{figure}
\centering
\includegraphics[width=0.44\textwidth]{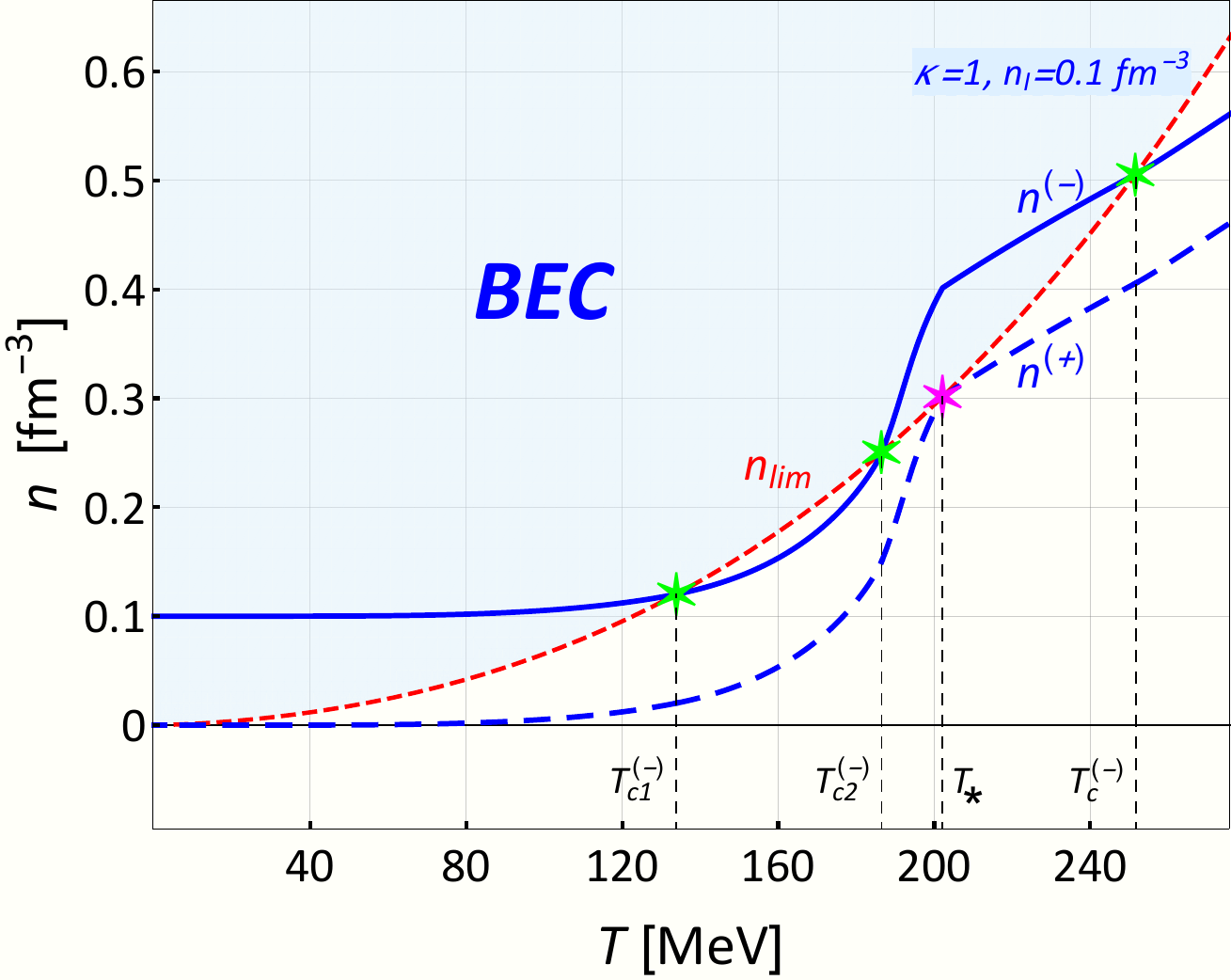}
\hspace{1mm}
\includegraphics[width=0.49\textwidth]{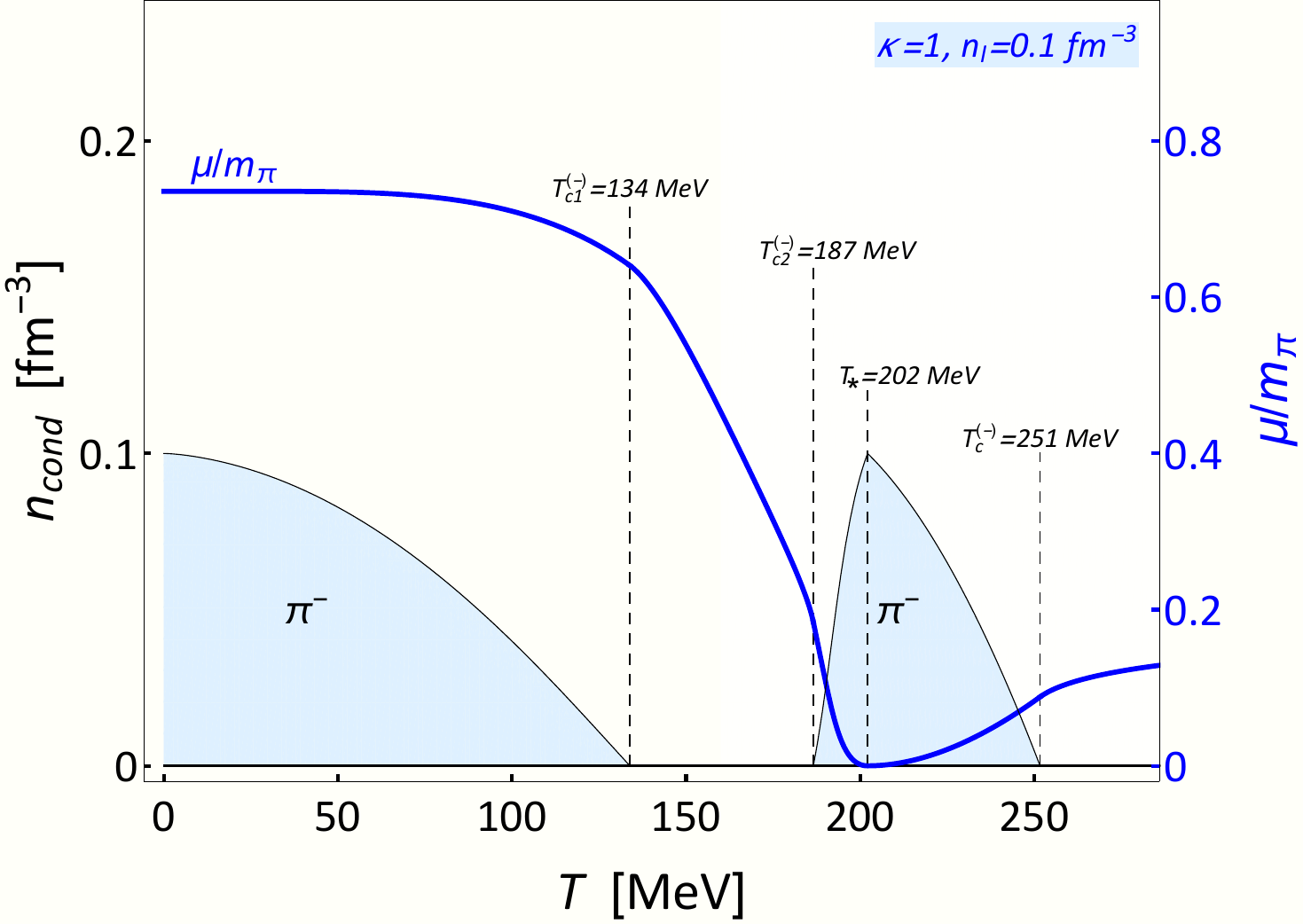}
\caption{ {\it Left panel:}
The particle-number densities $n^{(+)}$ , $n^{(-)}$ vs.
temperature for the interacting $\pi^+$-$\pi^-$ gas in the mean-field model
at $\kappa = 1$ and isospin density $n_I = 0.1\, \mathrm{fm}^{-3}$.
The temperature $T_{*}$ indicates the meta phase transition of the second order.
The temperatures $T_{\rm c}^{(-)}$, $T_{\rm c1}^{(-)}$ and $T_{\rm c2}^{(-)}$
indicate the multiple phase transitions of the second order
as in Fig.~\ref{fig:particles-antiparticles-kappa096_01}.
{\it Right panel:} The density of condensate (shaded region) vs. temperature and
the chemical-potential dependance on temperature (right Y-axis) for the same meson
system and the same conditions as in the left panel.
}
\label{fig:particles-antiparticles-kappa1_01}
\end{figure}

\begin{figure}
\centering
\includegraphics[width=0.49\textwidth]{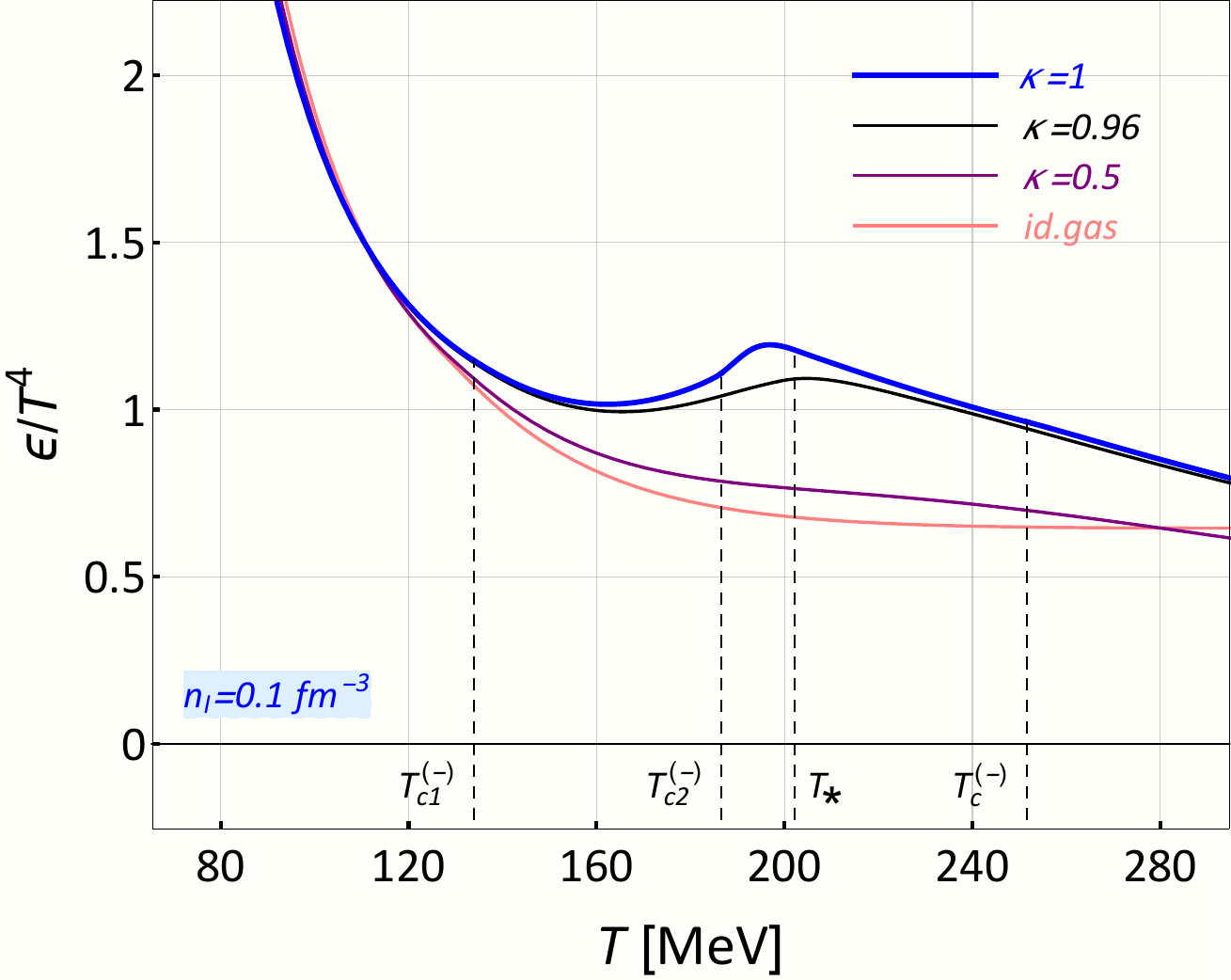}
\includegraphics[width=0.485\textwidth]{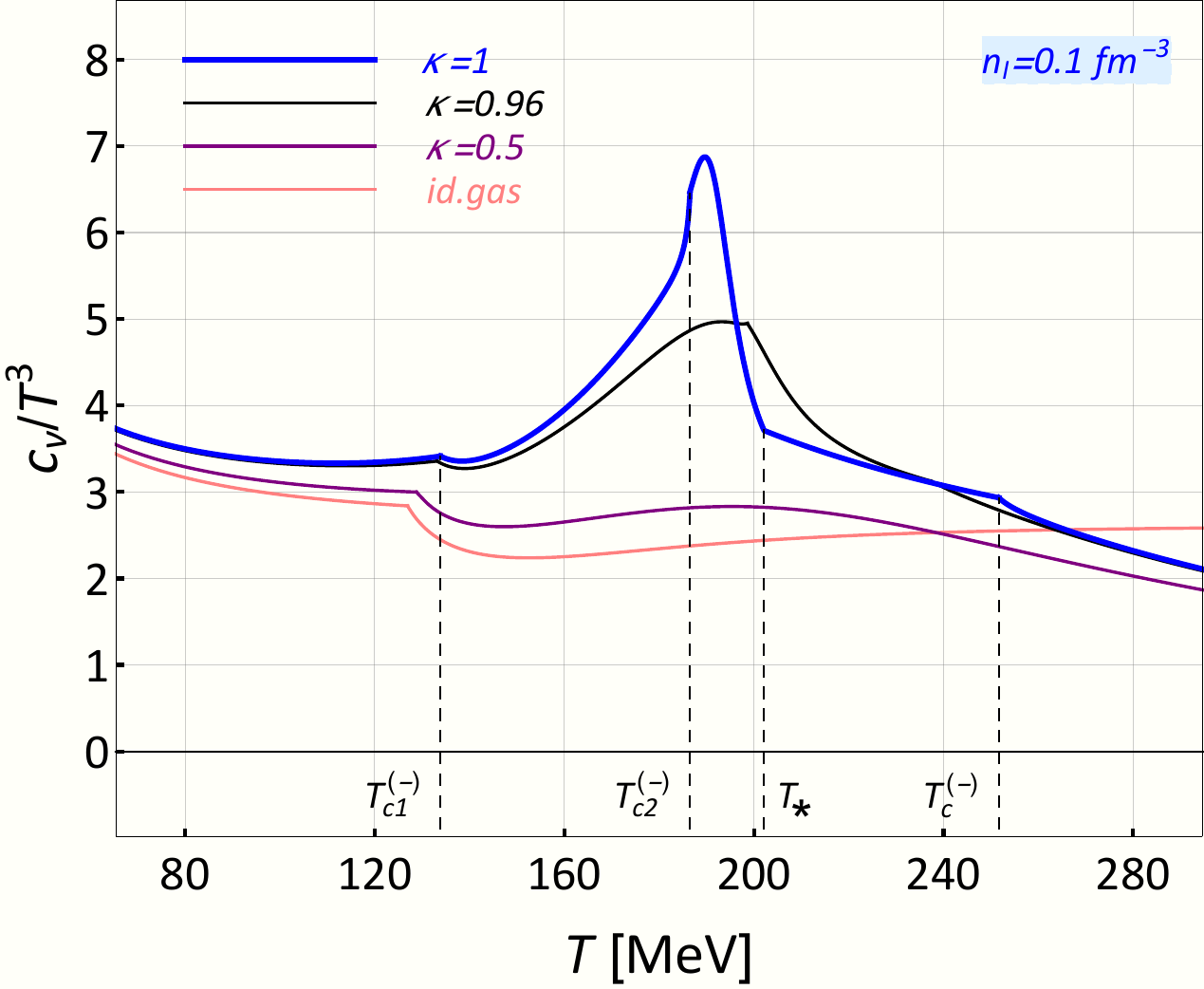}
\caption{
{\it Left panel:} Energy density vs. temperature at $n_{\rm I} = 0.1$~fm$^{-3}$
for the same particle-antiparticle system as in
Figs.~\ref{fig:particles-antiparticles-kappa1_01} at $\kappa = 0.5,\, 0.96,\, 1$.
The temperatures $T_{\rm c}^{(-)}$, $T_{\rm c1}^{(-)}$, $T_{\rm c2}^{(-)}$
and $T_{*}$ are the same as in Figs.~\ref{fig:particles-antiparticles-kappa1_01}.
{\it Right panel:}
Heat capacity vs. temperature for the same boson system and the same conditions
as in the left panel.
}
\label{fig:heat-capacity-kappa1-01}
\end{figure}

When the isospin density is finite, with increasing attraction between
particles, we arrive at the same specific metaphase transition due to the
behavior of the $\pi^+$ component.
Consider a system with a finite isospin density at $\kappa = 1$.
The left panel of Fig.~\ref{fig:particles-antiparticles-kappa1_01} shows the
temperature dependence of the density of antiparticles, i.e. $\pi^+$ mesons,
as a blue dashed line for the boson system at $n_I = 0.1$~fm$^{-3}$.
This is an interesting case, since there is only one point where the line
$n^{(+)}(T)$ touches the critical curve $n_{\rm lim}$; this point is marked
with a red asterisk and corresponds to the temperature $T_*$.
At this temperature, antiparticles experience a meta-like phase transition
of the second kind, since at this point:
a) the temperature dependence of the energy density is a smooth function,
see Fig.~\ref{fig:heat-capacity-kappa1-01}, left panel;
b) the temperature dependence of the heat capacity is a continuous function,
but has a break at temperature $T_*$, see Fig.~\ref{fig:heat-capacity-kappa1-01},
right panel;
c) there is no transition of antiparticles to a new phase, or no formation of
a condensate of $\pi^+$ mesons, i.e. the order parameter is zero on both sides
of this critical temperature $T = T_*$.
Properties (a) and (b) are formal evidence of a phase transition of the second
order.
Thus, according to calculations (a) and (b), the temperature $T_*$ indicates
a phase transition of the second order, but is not the starting point for the
formation of the order parameter (condensate).
Strictly speaking, when the temperature is lowered, there is no transition to
the condensate phase of $\pi^+$ mesons at the point $T_*$.
Therefore, we call this third type of phase transition, as in the case of
$n_I = 0$, a {\it meta phase transition of the second order} or a virtual phase
transition.

The right panel in Fig.~\ref{fig:particles-antiparticles-kappa1_01} shows the
formation of multiple condensates of particles, i.e. $\pi^-$ mesons.
The temperature dependence of the chemical potential is shown in the same panel
by a blue solid line (the right Y axis digitizes $\mu / m_{\pi}$).
Only at the point $T = T_{*}$ the necessary condition for the formation of the
condensate must be fulfilled simultaneously for both components of the boson
system:
1) for particles $m + U(n) - \mu_I = 0$;
2) for antiparticles $m + U(n) + \mu_I = 0$.
Therefore, at this point, two equations that come from the two necessary
conditions must be fulfilled: $m + U(n) = 0$ and $\mu_I = 0$.
Indeed, we see in the right panel of
Fig.~\ref{fig:particles-antiparticles-kappa1_01} that at temperature $T_{*}$
the chemical potential becomes zero (for details, see
\cite{Anch-PRC105,universe-2023}).

For the three types of phase transitions considered above in sections
\ref{sec:first-type}, \ref{sec:second-type} and \ref{sec:third-type},
the features of the behavior of the energy density and the heat capacity of
the interacting bosonic system are summarized in the left and right panels
of Fig.~\ref{fig:heat-capacity-kappa1-01}, respectively.
Indeed, it is clear that the appearance of new types of phase transitions is
due to an increase in the intensity of attraction between particles
(see the two upper curves for $\kappa = 0.96$ and $\kappa = 1$ in the right panel
of Fig.~\ref{fig:heat-capacity-kappa1-01}).


\section{The fourth type: Formation of a Bose condensate at finite temperature
$T < T_{\rm c}$}

\begin{figure}
\centering            
\includegraphics[width=0.47\textwidth]{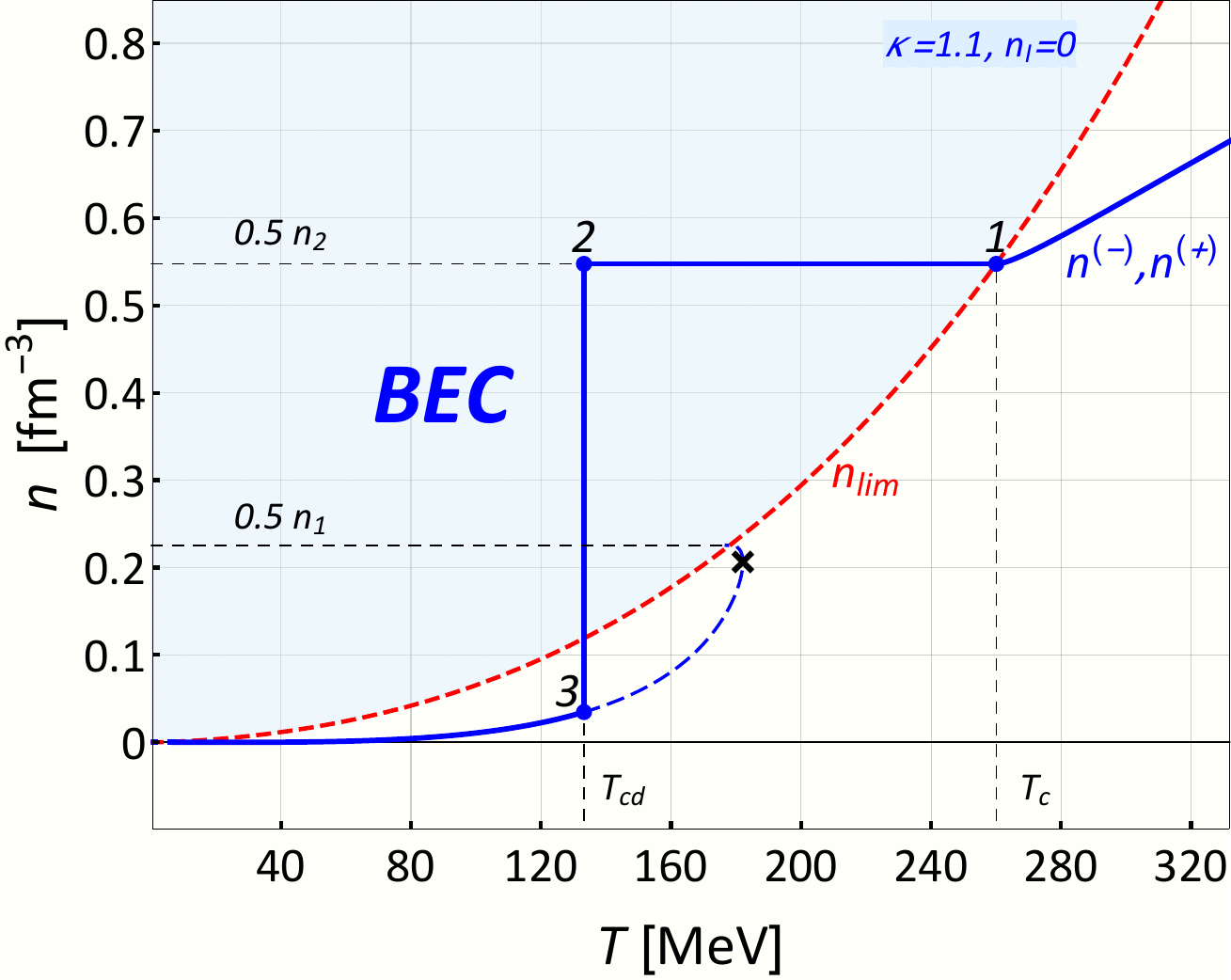}
\hspace{3mm}
\includegraphics[width=0.47\textwidth]{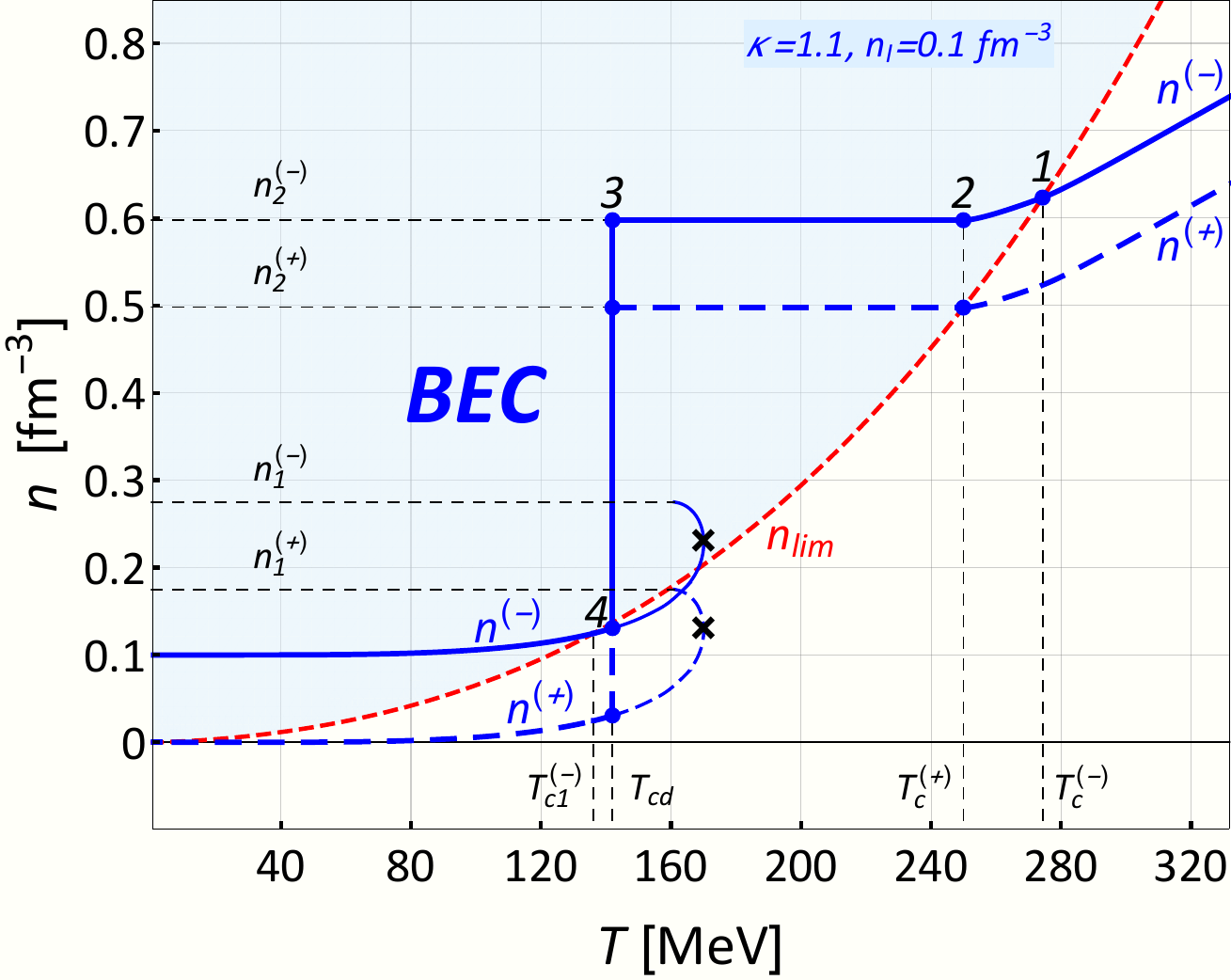}
\caption{
{\it Left panel:}
Temperature dependence of particle-number density $n^{(+)}$, $n^{(-)}$ for an
interacting $\pi^+$-$\pi^-$ pion gas in the mean field model.
The total isospin density is zero, $n_I = 0$.
The attraction parameter $\kappa = 1.1$.
{\it Right panel:}
Temperature dependence of particle-number density $n^{(+)}$, $n^{(-)}$ for an
interacting $\pi^+$-$\pi^-$ pion gas in the mean field model
at $n_I = 0.1\, \mathrm{fm}^{-3}$ and $\kappa = 1.1$.
}
\label{fig:kappa11_01}
\end{figure}

\begin{figure}
\centering
\includegraphics[width=0.45\textwidth]{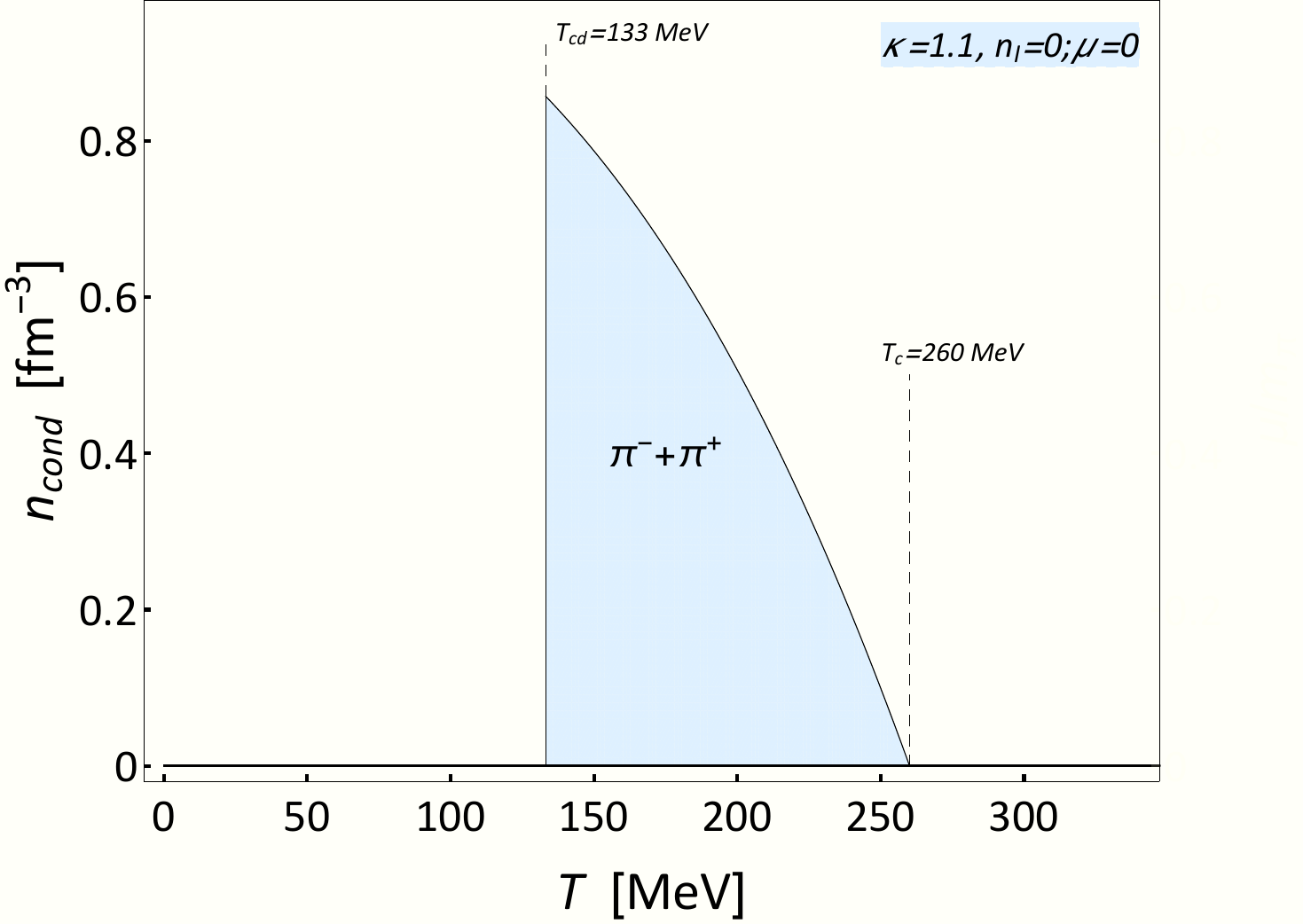}
\hspace{3mm}
\includegraphics[width=0.49\textwidth]{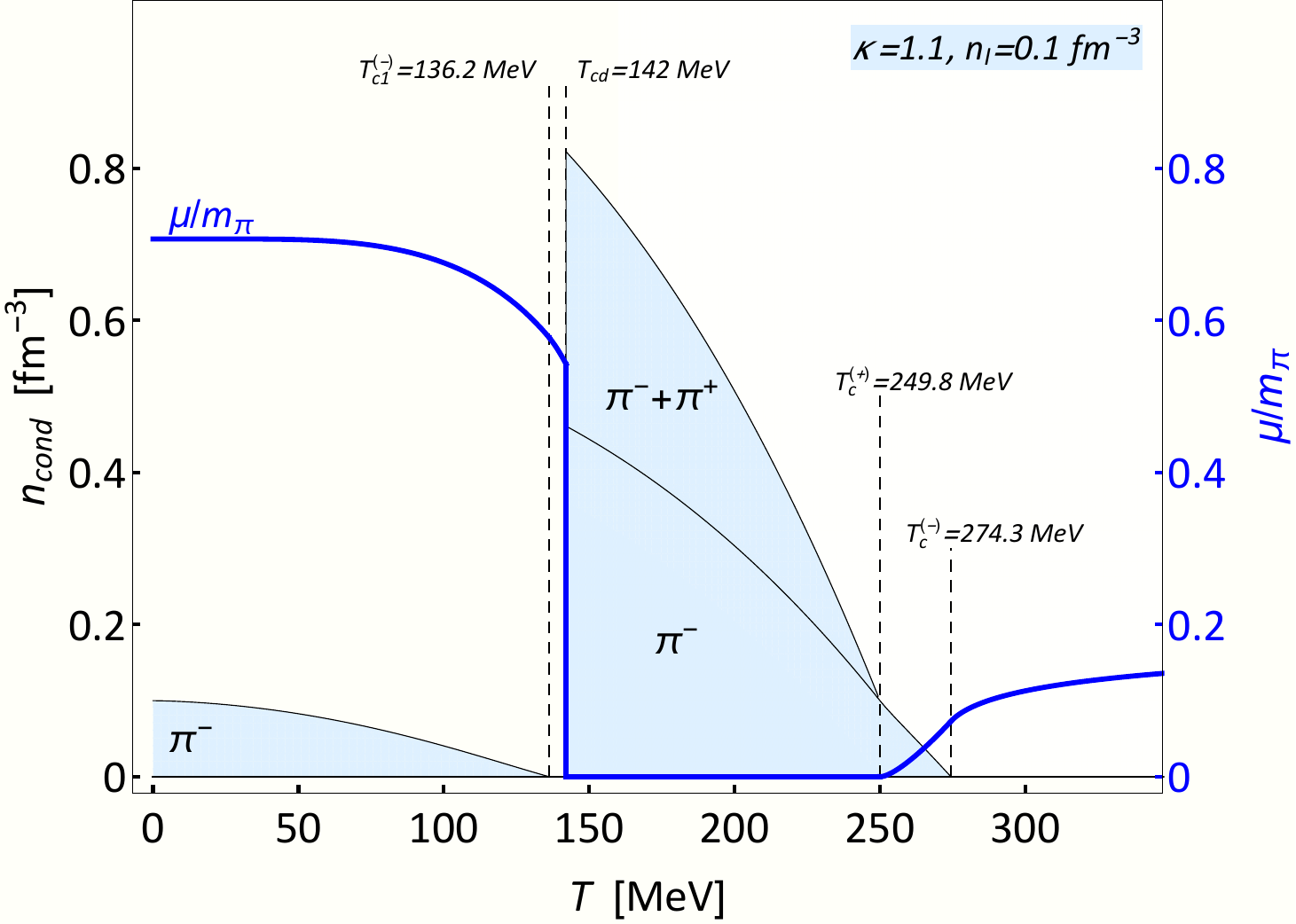}
\caption{ {\it Left panel:}
Dependence of the density of condensed particles on temperature at $n_I = 0$
and the attraction parameter $\kappa = 1.1$ for the same boson system and the
same conditions as in the left panel of Fig.~\ref{fig:kappa11_01}.
{\it Right panel:}
Dependence of the density of condensed particles on temperature and dependence
of the chemical potential (right Y axis) on temperature at
$n_I = 0.1\, \mathrm{fm}^{-3}$ and $\kappa = 1.1$ for the same boson system and
the same conditions as in the right panel of Fig.~\ref{fig:kappa11_01}.
}
\label{fig:condensate-kappa11_01}
\end{figure}

When the attraction becomes over-critical, i.e. $\kappa > 1$, the situation is
getting more complicated.
In this case
the results of the calculations of the number densities $n^{(-)}(T)$ and
$n^{(+)}(T)$ at $n_I = 0$, i.e. for the neutral two-component bosonic system,
are shown in the left panel of Fig.~\ref{fig:kappa11_01} for $\kappa = 1.1$.
At zero isospin (charge) density the system acts similar to single-component
system with degeneracy factor $g = 2$.
For the temperatures less than $T_{\rm cd}$ both components are in thermal
phase, but the system undergoes a first-order phase transition at a finite
temperature, $T = T_{\rm cd}$, leading to the formation of a condensate.
As energy is pumped into the system, the temperature remains constant at
$T_{\rm cd}$ because all release of energy is spent on creation of the
particle-antiparticle pairs until the total density
$n$ reaches the value $n = n_2$ that is solution of equation
\begin{equation}
m + U(n) \,=\, 0 \,,
\label{eq:gap}
\end{equation}
which is the necessary condition for the condensate formation
(remind, in the neutral system $\mu_I = 0$).
For example, for the mean field $U(n) = - An + Bn^2$, where coefficients $A$
and $B$ are positive, there are two roots of equation (\ref{eq:gap}):
$n_{1,2} = \sqrt{m/B} \left( \kappa \mp \sqrt{\kappa^2 - 1}\right)$,
where $\kappa  \equiv A/(2\sqrt{m B})$ and $n_1 \le n_2$.
It is evident that from two condensate branches $n_1$ and $n_2$ the higher
pressure has the branch with higher particle-number density, i.e. $n_2$.
Therefore, it is the second self-consistent branch that competes with the
pure thermal solution of the self-consistent equation (gap equation),
see for details Ref.~\cite{Anch-JPG46}.
The thermodynamic states of the system belonging to the $n = n_2$ branch have
a higher pressure than the thermal branch for temperatures $T > T_{\rm cd}$
(in the left panel of Fig.~\ref{fig:kappa11_01} the meta-stable states of the
thermal branch are shown as a blue dashed line for temperatures $T > T_{\rm cd}$).
After the first-order phase transition at $T = T_{\rm cd}$, with further energy
pumping, the bosonic system evolves along a constant density
$n^{(-)} = n^{(+)} = 0.5 n_2$ to the point $T = T_{\rm c}$, where the
second-order phase transition occurs.
That is, condensate appears in the system at the finite temperature
$T = T_{\rm cd}$ as a result of the first-order phase transition.
The system then evolves to a second-order phase transition at $T_{\rm c}$,
where the condensate disappears, see the left panel of
Fig.~\ref{fig:condensate-kappa11_01}.
We call this {\it fourth type} phase transition.

For finite isospin (charge) density and the over-critical attraction the
resulting picture is a little complicated.
The temperature dependence of the densities splits in such a way that
$n^{(-)}(T) - n^{(+)}(T) = n_I =$~const.
The necessary condition for the formation of the condensate must be
fulfilled simultaneously for both components of the boson system:
1) for particles $m + U(n) - \mu_I = 0$;
2) for antiparticles $m + U(n) + \mu_I = 0$.
Hence, in the condensate phase we get two equations: $m + U(n) = 0$ and
$\mu_I = 0$.
That is, in the bosonic system with $n_I \ne 0$ in the condensate phase
equations are the same as in the previous case for the neutral system with
$\mu_I = 0$.
In fact, there is no contradiction.
It must be recognized that the chemical potential governs the thermal fraction
of particles and antiparticles.
These two thermal fractions are equal to each other and $n_{\rm lim}$ when both
particles and antiparticles are in the condensate phase at the same time; see
the left panel of Fig.~\ref{fig:kappa11_01}.
The equality of the thermal particle density to the thermal antiparticle density
will be guaranteed when the chemical potential in the Bose-Einstein distribution
is zero; however, as we have seen at the same time, $U(n) = - m$ for both components.
Therefore, the thermal densities of both components are $n_{\rm lim}(T)$ in the
condensate phase.
An interesting point follows from this reasoning: the total charge of the
thermal fractions of particles and antiparticles is zero when the bosonic system
is in the condensate phase; the isospin charge is accumulated in the condensate
fraction, i.e. the total momentum of the additional charge is zero.
The temperature dependence of condensates is shown in the right panel of
Fig.~\ref{fig:condensate-kappa11_01} as the blue shadowed regions.
Here, the behavior of the chemical potential is shown by a blue solid line
(the value of the number is fixed on the right Y-axis).
It can be seen that indeed in the temperature interval
$T_{\rm cd} \le T \le T_{\rm c}^{(+)}$, where both $\pi^-$ and $\pi^+$ mesons
are in the condensate phase, the value of the chemical potential is zero.
It is also seen that the critical temperature $T_{\rm c}$ splits into two
$T_{\rm c}^{(-)}$ and $T_{\rm c}^{(+)}$, which corresponds to the separate exit
from the condensate phase of $\pi^-$ mesons and $\pi^+$ mesons.


\section{Conclusions}
We have demonstrated that increasing the intensity of attraction in an
interacting bosonic system leads to several new types of phase transitions.
In addition to the standard second-order phase transition, three new types
emerge: multiple condensate formation, meta phase transitions without
condensate formation, and condensate formation at finite temperature
$T < T_{\rm c}$ through a first-order phase transition.
We also showed that particles and antiparticles can simultaneously be in a
condensate state only in the case of a supercritical intensity of attraction
between particles, which in the mean field model occurs when the attraction
coefficient $\kappa > 1$.

\section*{Acknowledgements}
The work of D.A., V.Gn. and D.Zh. was supported by the Simons Foundation and by
the Program "The structure and dynamics of statistical and quantum-field systems"
of the Department of Physics and Astronomy of the NAS of Ukraine.
This work is supported also by Department of target training of
Taras Shevchenko Kyiv National University and the NAS of Ukraine,
grant No. 6$\Phi$-2024.
I.M. thanks FIAS for support and hospitality.
H.St. thanks for support from the J. M. Eisenberg Professor Laureatus of the
Fachbereich Physik.


\end{document}